\documentclass[pra,amsfonts,amsmath,twocolumn,superscriptaddress,nofootinbib]{revtex4}

\usepackage{amssymb,color,paralist,amsmath,epsfig}
\usepackage{graphicx}
\usepackage{comment}
\usepackage{amsfonts}
\usepackage{relsize}
\usepackage{nicefrac,esint}
\usepackage{float}

\usepackage[colorlinks,citecolor=blue,linktoc=all,linkcolor=cyan]{hyperref}

\newcommand{\beq}{\begin{equation}}
\newcommand{\eeq}{\end{equation}}

\newcommand{\ber}{\begin{eqnarray}}
\newcommand{\eer}{\end{eqnarray}}

\begin{document}

\title{Two-photon exchange on the neutron and the hyperfine splitting}

\author{Oleksandr Tomalak}
\affiliation{Department of Physics and Astronomy, University of Kentucky, Lexington, KY 40506, USA}
\affiliation{Fermilab, Batavia, IL 60510, USA}
\affiliation{Institut f\"ur Kernphysik and PRISMA Cluster of Excellence, Johannes Gutenberg Universit\"at, Mainz, D-55099, Germany}

\date{\today}

\begin{abstract}
We calculate the contribution from the two-photon exchange on the neutron to the hyperfine splitting of S energy levels. We update the value of the neutron Zemach radius and estimate total recoil and polarizability corrections. The resulting two-photon exchange in electronic atoms exceeds by an order of magnitude the leading Zemach term and has different sign both in electronic and muonic hydrogen.
\end{abstract}

\maketitle

\begin{figure}[h]
\begin{center}%\centering
\includegraphics[width=0.23\textwidth]{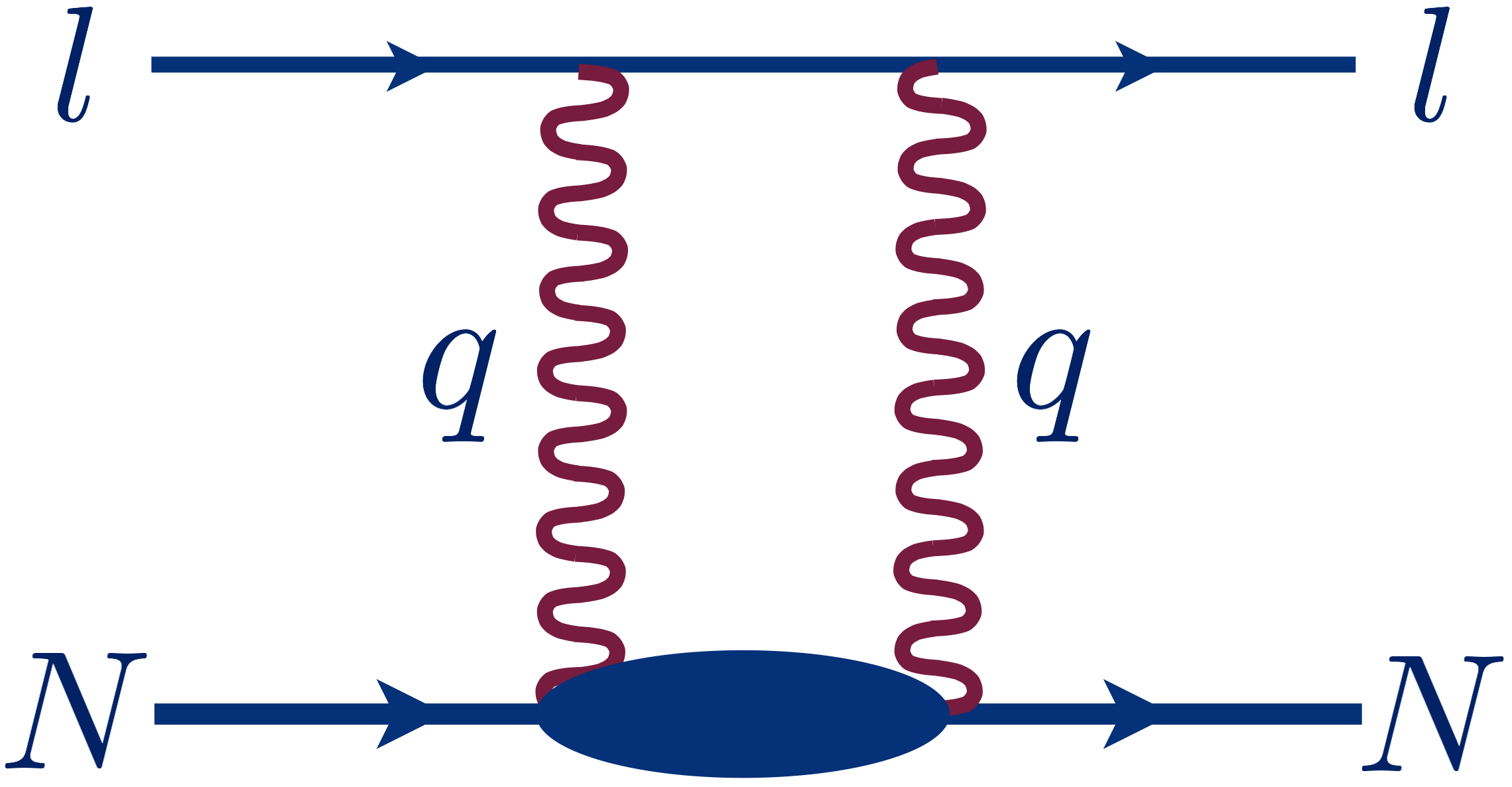}
\end{center}
\caption{Two-photon exchange graph. The contribution of the crossed graph is included into the lower blob.}
\label{TPE_graph}
\end{figure}

Modern spectroscopical measurements in light muonic atoms support the physics community with precise values of the Rydberg constant and nuclei electromagnetic radii \cite{Mohr:2015ccw,Tanabashi:2018oca,Yerokhin:2018gna}. The unexpected discrepancy between muonic and electronic values of the charge radius in hydrogen and deuterium \cite{Pohl:2010zza,Antognini:1900ns,Pohl1:2016xoo,Bernauer:2010wm,Bernauer:2013tpr,Mohr:2012tt} calls for revisiting the higher-order corrections with an emphasis on the uncertain hadronic and nuclei contributions. In particular, to analyze measurements of the hyperfine splitting in light muonic nuclei and to extract the precise value of the Zemach radius, the higher-order radiative corrections have to be taken into account \cite{Faustov:2014rea,Kalinowski:2018vjm}. In recent decades, the $O \left(\alpha^5 \right)$ contribution from the graph with two exchanged photons (TPE) on a proton and nucleus (see Fig. \ref{TPE_graph}) to the Lamb shift and hyperfine splitting was scrutinized by numerous authors \cite{Pachucki:1996zza,Faustov:1999ga,Pineda:2002as,Pineda:2004mx,Nevado:2007dd,Carlson:2011zd,Hill:2012rh,Birse:2012eb,Miller:2012ne,Alarcon:2013cba,Gorchtein:2013yga,Peset:2014jxa,Tomalak:2015hva,Caprini:2016wvy,Hill:2016bjv,Carlson:2008ke,Carlson:2011af,Eides:2000xc,Karshenboim:2005iy,Horbatsch:2016xx,Peset:2016wjq,Tomalak:2017lxo,Tomalak:2017npu,Tomalak:2017owk,Ji:2013oba,Hernandez:2014pwa,Dinur:2015vzv}. Besides the scattering on a proton, the TPE effect in light atoms contains contributions from nuclei excitations as well as from the scattering on a neutron. The contribution from the two-photon exchange on the neutron to the Lamb shift was recently investigated in Ref. \cite{Tomalak:2018uhr,NevoDinur:2018hdo}. For the hyperfine splitting, only the leading Zemach correction was evaluated in Refs. \cite{Friar:2005yv,Friar:2005je} from parametrizations of the neutron form factors.

In this work, we reproduce the result of Refs. \cite{Friar:2005yv,Friar:2005je} exploiting the modern form factor parametrizations which satisfy the consistency criterium of Ref. \cite{Tomalak:2017npu} for such a calculation. We account for the two-photon exchange effects beyond the Zemach term by means of the forward Compton scattering amplitudes \cite{Tomalak:2017owk}. We estimate the polarizability contribution from the MAID partial-wave solution and illustrate how good such an estimate can be on the example of the TPE on the proton.
 
We determine the hyperfine-splitting correction from the forward scattering amplitude at threshold. It is convenient to express the resulting hyperfine splitting and individual contributions in terms of the effective radii. The resulting nucleon radius $r^\mathrm{N}_{2\gamma}$ is given as a sum of three terms:
\ber
r^\mathrm{N}_{2\gamma} = r_\mathrm{Z} + r_\mathrm{R} + r_\mathrm{pol},
\eer
where $r_\mathrm{Z}$, $r_\mathrm{R}$ and $r_\mathrm{pol}$ stand for the Zemach, recoil and polarizability radii, respectively (the terminology is taken from the hydrogen TPE). These radii are expressed as the photon energy $\nu_\gamma$ and virtuality $Q^2$ integrals over the neutron electric $\mathrm{G}_\mathrm{E}$ and magnetic $\mathrm{G}_\mathrm{M}$ form factors and polarized spin structure functions $g_1,~g_2$ \cite{Iddings:1959zz,Drell:1966kk,Eides:2000xc,Carlson:2008ke,Tomalak:2017owk}:
\ber
 r_\mathrm{Z} &=& - \frac{2}{\pi }  \int \limits^{\infty}_{0} \frac{\mathrm{d} Q^2}{Q^3}  \frac{\mathrm{G}_\mathrm{E}\left(Q^2\right) \mathrm{G}_\mathrm{M}\left(Q^2\right)  }{\mathrm{\mu}_\mathrm{N}}, \\
r_\mathrm{R}  &=&-\frac{1}{2} \int \limits^{\infty}_{0} \frac{\mathrm{d} Q^2}{Q^2}  \frac{ \left( 2 + \rho\left(\tau_\mathrm{l}\right) \rho\left(\tau_\mathrm{N} \right)  \right) \mathrm{F}_1\left(Q^2\right)  }{ \sqrt{\tau_\mathrm{N}} \sqrt{1+\tau_\mathrm{l}} + \sqrt{\tau_\mathrm{l}} \sqrt{1+\tau_\mathrm{N}} } \frac{ \mathrm{G}_\mathrm{M}\left(Q^2\right)  }{ \pi \mathrm{\mu}_\mathrm{N} m_r} \nonumber \\ 
&-&  \frac{3}{2} \int \limits^{\infty}_{0} \frac{\mathrm{d} Q^2}{Q^2}  \frac{ \rho\left(\tau_\mathrm{l}\right) \rho\left(\tau_\mathrm{N} \right) \mathrm{F}_2\left(Q^2\right)  }{ \sqrt{\tau_\mathrm{N}} \sqrt{1+\tau_\mathrm{l}} + \sqrt{\tau_\mathrm{l}} \sqrt{1+\tau_\mathrm{N}} } \frac{ \mathrm{G}_\mathrm{M}\left(Q^2\right)  }{\pi \mathrm{\mu}_\mathrm{N} m_r}  \nonumber \\
&-& r_\mathrm{\mathrm{F}_2^2} - r_\mathrm{Z} ,  \label{recoil_correction} \\
r_\mathrm{pol}\hspace{-0.05cm} &=& \hspace{-0.05cm}- \int \limits^{\infty}_{0} \frac{\mathrm{d} Q^2}{Q^2} \int \limits^{\infty}_{\nu^{\mathrm{inel}}_{\mathrm{thr}}} \frac{\mathrm{d} \nu_\gamma}{\nu_\gamma}  \frac{\left( 2 + \rho\left(\tau_\mathrm{l}\right) \rho\left(\tilde{\tau}\right) \right)  \frac{g_1 \left(\nu_\gamma, Q^2 \right)}{\pi \mathrm{\mu}_\mathrm{N}m_r}}{  \sqrt{\tilde{\tau}} \sqrt{1+\tau_\mathrm{l}} + \sqrt{\tau_\mathrm{l}} \sqrt{1+\tilde{\tau}}   } \nonumber \\
 &+& 3 \int \limits^{\infty}_{0} \frac{\mathrm{d} Q^2}{Q^2} \int \limits^{\infty}_{\nu^{\mathrm{inel}}_{\mathrm{thr}}} \frac{\mathrm{d} \nu_\gamma}{ \nu_\gamma} \frac{1}{\tilde{\tau}}  \frac{ \rho\left(\tau_\mathrm{l}\right) \rho\left(\tilde{\tau} \right)  \frac{g_2 \left(\nu_\gamma, Q^2 \right)}{\pi \mathrm{\mu}_\mathrm{N}m_r} }{  \sqrt{\tilde{\tau}} \sqrt{1+\tau_\mathrm{l}} + \sqrt{\tau_\mathrm{l}} \sqrt{1+\tilde{\tau}}   } \nonumber \\
&+& r_\mathrm{\mathrm{F}_2^2} , \label{polar_correction} \\
r_\mathrm{\mathrm{F}_2^2} &=& -\frac{m}{4 M} \int \limits^{\infty}_{0} \frac{\mathrm{d} Q^2}{Q^2}  \frac{\rho (\tau_l) \left( \rho (\tau_l) - 4 \right) \mathrm{F}^2_2 \left(Q^2\right)}{\pi \mu_\mathrm{N} m_r}, \label{F22_term}
\eer
with Pauli $ \mathrm{F}_1$ and Dirac $ \mathrm{F}_2$ form factors, lepton and nucleon masses $m$ and $M$, respectively, the reduced mass $m_r = M m / \left( M +m \right)$, the nucleon magnetic moment $\mu_\mathrm{N}$, the pion-nucleon inelastic threshold $ \nu^{\mathrm{inel}}_{\mathrm{thr}} $ (with pion mass $m_\pi$),
\ber
\nu^{\mathrm{inel}}_{\mathrm{thr}}  = m_\pi + \frac{m^2_\pi + Q^2}{2 M},
\eer and the following notations:
\ber
 \tau_\mathrm{l} &=& \frac{Q^2}{4 m^2},\qquad \tau_\mathrm{N} = \frac{Q^2}{4 M^2},\qquad  \tilde{\tau} = \frac{\nu_\gamma^2}{Q^2}, \\  \label{taus}
 \rho(\tau) &=& \tau - \sqrt{\tau ( 1 + \tau )}.
\eer

The resulting contribution to the hyperfine structure of S energy levels $\delta \mathrm{E}$ induced by individual nucleons is expressed as \cite{Kalinowski:2018vjm,Pachucki:2008zz}~\footnote{To avoid double counting, one has to be careful combining TPE with other corrections.}
\ber
\delta \mathrm{E} = - \frac{16}{3} \pi \alpha^2 \frac{\psi^2 \left( 0 \right)}{M + m}  \vec{s}_l <\sum \limits_\mathrm{N} \mu_\mathrm{N} \vec{s}_\mathrm{N} r^\mathrm{N}_{2\gamma}> ,
\eer
with the spin operators of nucleon $\vec{s}_\mathrm{N}$ and lepton $\vec{s}_l$, the atomic wave function at origin $\psi \left( 0 \right)$ and the fine structure constant $\alpha$.

The leading for the proton, Zemach correction is sensitive to low-$Q^2$ region of the electromagnetic form factors. In this region, it can be parametrized by relatively well-known electric and magnetic radii \cite{Karshenboim:2014maa,Karshenboim:2014vea,Tomalak:2017npu} up to some splitting value $Q_0^2$. For the neutron, the leading terms in such an expansion $r^\mathrm{LE}_\mathrm{Z}$ are given by%~\footnote{ For the proton with nonzero electric charge, the expansion is symmetric w.r.t. the electric and magnetic radii:%
\ber
 r^\mathrm{LE}_\mathrm{Z} \approx  \frac{2 r_\mathrm{E}^2}{3 \pi }  \int \limits^{Q_0}_{0} \mathrm{d} Q \left( 1 - \frac{r_\mathrm{M}^2 Q^2}{6}\right) = \frac{2 r_\mathrm{E}^2 Q_0}{3 \pi } - \frac{r_\mathrm{E}^2 r_\mathrm{M}^2 Q_0^3}{27 \pi}. \label{Zemach_in_terms_of_others}
\eer
In the numerical evaluation, we take the p.d.g. values for the electric $r_\mathrm{E}$ and magnetic $r_\mathrm{M}$ radii \cite{Tanabashi:2018oca}:
\ber
r^2_\mathrm{E} &=& -0.1161\pm 0.0022~\mathrm{fm}^2, \\
r_\mathrm{M} &=& 0.864 \pm 0.009~\mathrm{fm}.
\eer
We use form factor parametrizations above. The dependence of the Zemach correction on the splitting parameter $Q_0^2$ provides a consistency check in the evaluation of this contribution \cite{Tomalak:2017npu}. At low $Q_0^2$, it has to show a plateau behavior for form factor fits which are in agreement with electric and magnetic radii since one can use the low momentum transfer expansion of form factors or fits themselves. In the following Fig. \ref{consistency}, we present the dependence of the Zemach correction on the splitting parameter $Q_0^2$ for different form factor parametrizations available in the literature. While the value of magnetic radius differs between fits less than by $5\%$, the electric radius varies significantly. For instance, the fit of Eq. (43) in Ref. \cite{Punjabi:2015bba} incorporates half the squared electric charge radius and the fit of Eq. (31) in Ref. \cite{Punjabi:2015bba} favors the squared charge radius that is 3 times larger than the p.d.g. value. It is not surprising that the fit of Ref. \cite{Ye:2017gyb} perfectly passes the consistency criterium since this fit was constrained by radii values.
\begin{figure}[h]
\begin{center}%\centering
\includegraphics[width=0.5\textwidth]{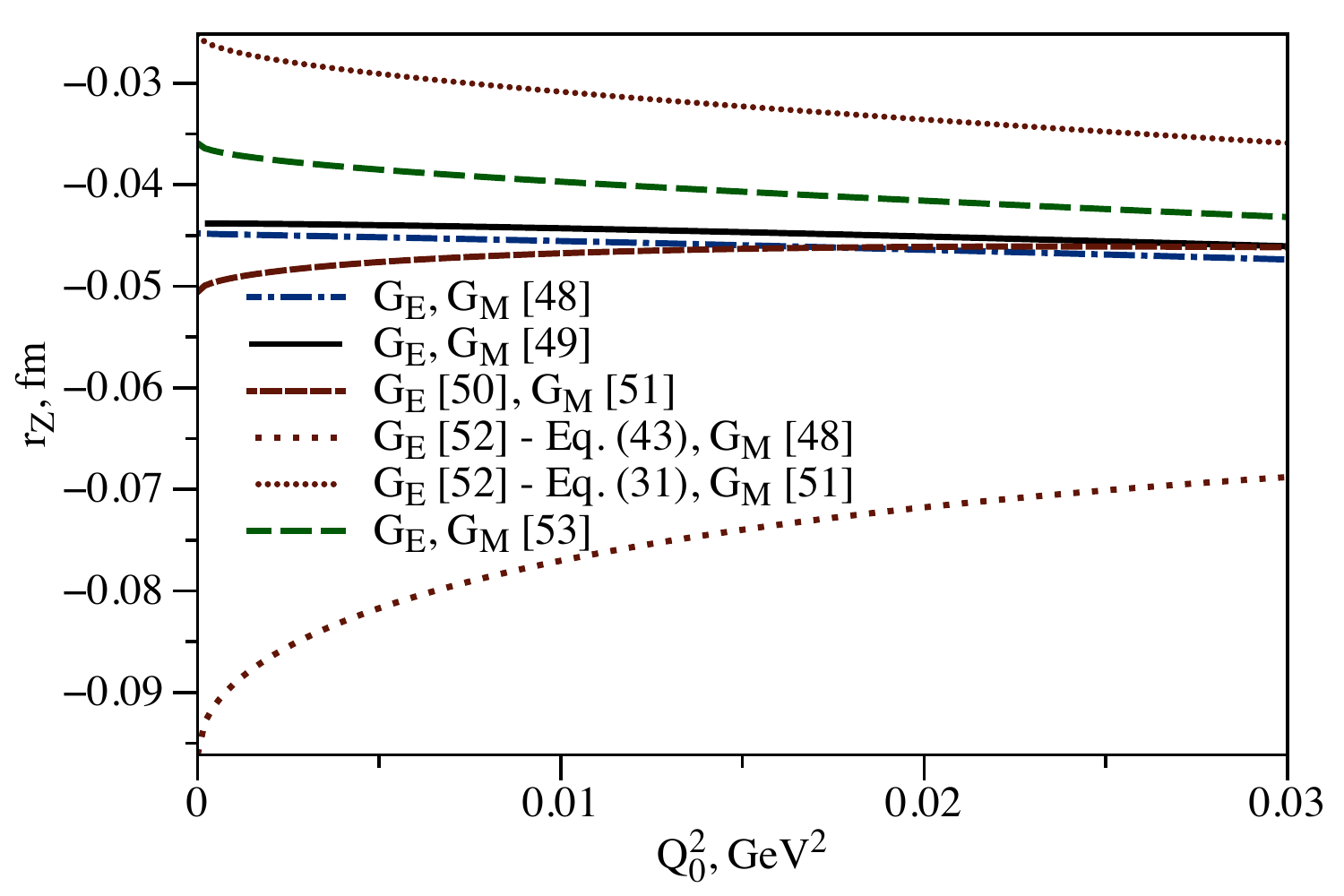}
\end{center}
\caption{Zemach correction as a function of the splitting parameter $Q^2_0$ into the low and high momentum transfer regions for different form factor parametrizations. The low-$Q^2$ contribution is saturated mainly by the charge and magnetic radii terms of Eq. (\ref{Zemach_in_terms_of_others}) for shown $Q^2_0$ values. Consistent with electromagnetic radii form factors represent a plateau behavior at low $Q^2_0$. The lowest curve is based on the electric form factor with 3 times larger squared charge radius, while the upper curve corresponds to the form factor with half the squared radius compared to the p.d.g. value }
\label{consistency}
\end{figure}
Evaluating the Zemach correction, we exploit form factor parametrizations which produce only small deviations at very low $Q_0^2$, i.e., fits of Refs. \cite{Kelly:2004hm,Ye:2017gyb}. Such parametrizations incorporate consistent values of electric and magnetic radii. We set $Q_0^2 = 0.01~\mathrm{GeV}^2$, when we can safely neglect contributions from higher moments of form factors expansion, and average over these fits. As an uncertainty estimate, we add the difference between two fits and the error due to the variation of the splitting parameter in the range $0.005~\mathrm{GeV}^2 \lesssim Q^2_0 \lesssim 0.02~\mathrm{GeV}^2$ in quadrature.  

For other contributions, we obtain the central value averaging over all form factor parametrizations from Refs. \cite{Kubon:2001rj,Warren:2003ma,Kelly:2004hm,Bradford:2006yz,Punjabi:2015bba,Ye:2017gyb} and estimate the uncertainty as a difference between the largest and smallest results. Our contributions for the neutron state are summarized in Table \ref{results_neutron}.
\begin{table}[h] 
\begin{center}
\begin{tabular}{|c|c|c|c|}
\hline
 $\mathrm{fm}$ &   $e n$ & $\mu n$  \\ \hline
$ r_\mathrm{Z}$ &   -0.0449 (13) &-0.0449 (13)  \\ 
$ r_\mathrm{R}$  &   0.328 (2) &0.0823 (8)  \\ 
$ r_\mathrm{Z} + r_\mathrm{R}$  &   0.284 (2) &  0.0374 (15)\\ 
$ r_\mathrm{\mathrm{F}_2^2}$ &  0.987 (7) & 0.260 (6) \\     \hline 
%$r_{\mathrm{F^2_2}}$ and low-$Q$ $\mathrm{I}_1$  &0.021 & 0.021 \\
$r_{\mathrm{F^2_2}}$ and low-$Q$ $\mathrm{I}_1$  &0.064 (32)  & 0.065 (32) \\
MAID ($\pi N$ only)  &  -0.012  & -0.009 \\ 
MAID ($\pi N$)  &  0.010  & 0.013 \\ 
MAID ($\pi N,~K N,~\eta N$)  &  0.015  & 0.018 \\ \hline
$ r_\mathrm{pol}$ & 0.064 (38)  & 0.065 (39) \\     \hline 
$ r^n_{2 \gamma}$ & 0.347 (38)& 0.102 (39) \\     \hline 
%$ r_\mathrm{pol}$ & 0.021 (13)  & 0.021 (13)  \\     \hline 
%$ r^n_{2 \gamma}$ & 0.31 (1)& 0.06 (1) \\     \hline 
\end{tabular}
\caption{Effective neutron-structure radii contributing to the hyperfine splitting in electronic and muonic atoms due to the neutron intermediate state. "$r_{\mathrm{F^2_2}}$ and low-$Q$ $\mathrm{I}_1$" is a sum of $\mathrm{I}_1 \left( 0 \right)' $ contribution from the low momentum transfer integration region and form factors contributions to $r_\mathrm{pol}$. The radius labeled "MAID ($\pi N$ only)" represents a pure correction from $\pi N$ states and does not include the result of the row "$r_{\mathrm{F^2_2}}$ and low-$Q$ $\mathrm{I}_1$", while "MAID ($\pi N$)" is a sum of two upper rows. "MAID ($\pi N,~K N,~\eta N$)" represents the contribution of $\pi N,~K N,~\eta N$ states together with "$r_{\mathrm{F^2_2}}$ and low-$Q$". Based on the similar to the proton saturation pattern, the resulting polarizability correction is estimated as "$r_{\mathrm{F^2_2}}$ and low-$Q$ $\mathrm{I}_1$" value. The error from the neutron intermediate state represents mainly the uncertainty due to the difference in fits central values and is presumably underestimated.} \label{results_neutron}
\end{center}
\end{table}
The resulting radius $r_\mathrm{Z} + r_\mathrm{R}$ is larger than the Zemach term and has an opposite sign. The term $r_\mathrm{R}$ saturates mainly at scales of the lepton mass and can be roughly estimated in the leading logarithmic approximation for pointlike nucleons, see Refs. \cite{Arnowitt:1953zz,GROTCH:1969zz,Khriplovich:2003nd}. It is remarkable that the piece subtracted from the elastic correction $ r_\mathrm{\mathrm{F}_2^2}$ exceeds all other terms.

The Zemach radius of the neutron $r_\mathrm{Z} $ is much smaller than the proton Zemach radius $r^\mathrm{p}_\mathrm{Z} \sim 1.06~\mathrm{fm}  $ \cite{Tomalak:2017npu} and has an opposite sign. The origin of this difference is in the overall zero electric charge of the neutron. In contrast to the proton, the correction $r_\mathrm{R}$ differs significantly between muonic and electronic atoms. The integrand of this correction has a definite sign in case of the neutron, while changes sign in electronic hydrogen. This sign change between electron and hadron mass scales results into similar values of $r_\mathrm{R}$ for the proton in electronic and muonic atoms.

For the polarizability correction, we replace the first moment of the $g_1$ structure function $\mathrm{I}_1 \left( Q^2 \right)$:
\ber
\mathrm{I}_1\left( Q^2 \right) & =& \int \limits^{\infty}_{ \nu^{\mathrm{inel}}_{\mathrm{thr}}} g_1 \left(\nu_\gamma, ~Q^2\right) \frac{M \mathrm{d} \nu_\gamma}{\nu_\gamma^2}, \label{GDH_type}
\eer
by the Gerasimov-Drell-Hearn (GDH) sum rule \cite{Drell:1966jv,Gerasimov:1965et}. More specifically, we add the term given by the replacement $\mathrm{F}^2_2 \left(Q^2\right) \to 4 \mathrm{I}_1\left( Q^2 \right)$ in Eq. (\ref{F22_term}) \cite{Tomalak:2017owk} and subtract the same term expressed through the spin structure functions with Eq. (\ref{GDH_type}). The latter is evaluated from the polarized spin structure functions together with other $g_1,g_2$-dependent pieces of $r_\mathrm{pol}$. For the contribution from $ 4 \mathrm{I}_1\left( Q^2 \right) + \mathrm{F}^2_2 \left(Q^2\right)$, we use the expansion $ \mathrm{I}_1\left( Q^2 \right)  \approx  \mathrm{I}_1\left( 0 \right) + \mathrm{I}_1\left( 0 \right)' Q^2 $ at the low momentum transfer region and connect it to the data-based integrand at high $Q^2$. 

For the neutron,
\ber
\mathrm{I}_1(0) & = & - \frac{\mathrm{\mu}_\mathrm{n}^2}{4}, \label{I1_moment}
\eer
and the derivative term $\mathrm{I}_1 \left( 0 \right)' \approx 6 ~\mathrm{GeV}^{-2}$ which is estimated to be slightly smaller than for the proton \cite{Prok:2008ev} since the data of Ref. \cite{Deur:2004ti} and ChPT calculations in Ref. \cite{Lensky:2014dda} indicate a slightly positive slope of the proton-neutron difference in $\mathrm{I}_1 \left( Q^2 \right)$. For the evaluation of $\mathrm{I}_1$ above $Q \gtrsim 0.2-0.25~\mathrm{GeV}$ and in other parts of the calculation, we use the MAID parametrization as an input \cite{Drechsel:1998hk,Drechsel:2007if} and sum over the $\pi N,~K N$ and $\eta N$ channels. We add MAID contributions on top of the low-$Q^2$ behavior of $\mathrm{I}_1$ and $r_{\mathrm{F^2_2}}$; see Table \ref{results_neutron} for details. We are not able to evaluate the resulting polarizability radius $r_\mathrm{pol}$ directly from the data due to the lack of polarized spin structure functions for other intermediate states.

To test such evaluation and estimate the polarizability correction, we compare the results from the structure functions measurements \cite{Kuhn:2008sy,Sato:2016tuz,Fersch:2017qrq} to the MAID-based evaluation of HFS on the proton in Table \ref{consistency}.  We notice that the effect of $~K N$ and $\eta N$ channels is much smaller than the leading $\pi N$ contribution as in case of the neutron.
\begin{table}[h] 
\begin{center}
\begin{tabular}{|c|c|c|c|}
\hline
 $\mathrm{fm}$ &   $e p$ & $\mu p$  \\ \hline
$ r_\mathrm{Z}$  \cite{Tomalak:2017npu,Tomalak:2018uhr} &   1.055 (13) &1.055 (13)  \\ 
$ r_\mathrm{R}$ \cite{Tomalak:2017npu,Tomalak:2018uhr}  &   -0.1411 (14) &-0.1203 (8)  \\ 
$ r_\mathrm{Z} + r_\mathrm{R}$ \cite{Tomalak:2017npu,Tomalak:2018uhr} &   0.914 (13) &  0.935 (13)\\  
$ r_\mathrm{\mathrm{F}_2^2}$\cite{Tomalak:2017owk} &  -0.596 (2) & -0.158 (1) \\     \hline 
%$ r_\mathrm{pol}$ \cite{Tomalak:2018uhr}  &   -0.0509  &  -0.0518\\ 
%$r_{\mathrm{F^2_2}}$ and $I_1$ &-0.0493 & -0.0474 \\
%MAID ($\pi N$ only)  &  0.0269  & 0.0253 \\ 
%MAID ($\pi N$)& -0.0224 & -0.0221 \\     
%MAID ($\pi N,~K N,~\eta N$)& -0.0274 & -0.0273 \\     \hline 
$r_{\mathrm{F^2_2}}$ and low-$Q$ $\mathrm{I}_1$  &-0.049 & -0.047 \\
MAID ($\pi N$ only)  &  0.027  & 0.025 \\ 
MAID ($\pi N$)& -0.022 & -0.022 \\     
MAID ($\pi N,~K N,~\eta N$)& -0.027 & -0.027 \\     \hline 
$ r_\mathrm{pol}$ \cite{Tomalak:2017npu,Tomalak:2018uhr}  &   -0.051 (13) &  -0.052 (13)\\ \hline
$ r^{p}_\mathrm{2\gamma}$ from eH 1S HFS \cite{Tomalak:2017lxo,Tomalak:2018uhr}  &   0.861 (6)   &  0.880 (8)\\ \hline
$ r^{p}_\mathrm{2\gamma}$ \cite{Tomalak:2017npu,Tomalak:2018uhr}  &   0.863 (20)  & 0.883 (19)\\ \hline
\end{tabular}
\caption{Effective proton-structure radii contributing to the hyperfine splitting in electronic and muonic atoms. The values in upper four and lower three rows are taken from the corresponding references. Other rows represent the saturation pattern of the polarizability correction; see Table \ref{results_neutron} for description.} \label{results_proton}
\end{center}
\end{table}
As it was found in Ref. \cite{Tomalak:2017lxo}, the effective polarizability radii in electronic and muonic hydrogen based on structure functions of Refs. \cite{Kuhn:2008sy,Sato:2016tuz,Fersch:2017qrq} are close to each other.

%proton
%K1:  -0.000398   -0.000418
%K2:  -0.001329   -0.00141
%eta:  -0.00323   -0.00335
%piN GDH: -0.6522221745 from -0.8035754105
%K1 GDH: 0.00348
%K2 GDH: 0.00359
%eta GDH: 0.0350

%neutron
%K1:  0.000231   0.000246
%K2:  0.00205   0.00219
%eta:  0.003119   0.00323
%piN GDH: -0.5162633379 from -0.9149331121
%K1 GDH: 0.00317
%K2 GDH: 0.00815
%eta GDH: 0.0224
Likewise the GDH sum rule~\cite{Drechsel:2000ct,Drechsel:2004ki}, the saturation pattern of the contributions from different channels from the neutron is similar to the proton case with slightly different decomposition. This allows us to estimate the central value of the polarizability correction $r_\mathrm{pol}$ in Table \ref{results_neutron} roughly as the sum of the $r_{\mathrm{F^2_2}}$ term and the low-$Q^2$ part of $\mathrm{I}_1$. The polarizability correction for the neutron has the same order of magnitude as the Zemach term and an opposite sign. The effective polarizability radius for electronic and muonic atoms is of the same size. We add an error $60\%$ to $r_\mathrm{pol}$ and summarize the resulting HFS correction in Table \ref{results_neutron}.

We evaluated the hyperfine splitting correction from the two-photon exchange on the neutron and presented results in terms of effective radii which generalize the Zemach radius. Our full result is larger than the known Zemach radius of the neutron and has an opposite sign. In the case of electronic atoms, the total contribution is an order of magnitude larger than the Zemach term. As in hydrogen, the Zemach and polarizability radii are similar in electronic and muonic atoms, while the radius $r_\mathrm{R}$ differs significantly in case of the neutron. The obtained results will be useful in the evaluation of the structure corrections to the hyperfine splitting in light atoms and can be improved in a future with a progress in the understanding of the neutron form factors, with account for other intermediate states, especially $\pi \pi N$, and extractions of the neutron spin structure functions.

We acknowledge Krzysztof Pachucki for pointing on the missing input in the nuclear structure corrections to the hyperfine splitting. We acknowledge Krzysztof Pachucki and Richard Hill for careful reading of the manuscript and useful discussions. This work was supported in part by a NIST precision measurement grant and by the U. S. Department of Energy, Office of Science, Office of High Energy Physics, under Award No. DE-SC0019095. This work was partially supported by the Deutsche Forschungsgemeinschaft (DFG) through the Collaborative Research Center ``The Low-Energy Frontier of the Standard Model'' (SFB 1044). The author would like to acknowledge the Fermilab theory group for its hospitality and support.

\end{document}